\documentclass[article,twocolumn,floatfix,aps,pre,amsmath,showpacs]{revtex4}

\usepackage{graphicx}
\usepackage{amssymb}

\begin{document}

\title{Metabasin dynamics and local structure in supercooled water}

\def\roma{\affiliation {
Dipartimento di Fisica and CNR-INFM Udr and CRS-SOFT: Complex Dynamics in Structured Systems, Universit\`a
di Roma ``La Sapienza'', Piazzale Aldo Moro 2, I-00185, Roma,
Italy}}
\def\argentina{\affiliation{
Fisicoqu\'{\i}mica, Departamento de Qu\'{\i}mica, Universidad Nacional del Sur, Av. Alem 1253,
8000 Bah\'{\i}a  Blanca, Argentina.}
}
\author{Jorge Ariel Rodr\'{\i}guez Fris}\argentina 
\author{and Gustavo A. Appignanesi}\argentina

\author{Emilia La Nave}\roma
\author{Francesco Sciortino}\roma

\date{\today}

\begin{abstract}

We employ the Distance Matrix method to investigate metabasin dynamics in supercooled water. 
We find that the motion of the system consists in the exploration of a finite region of configuration space
(enclosing several distinct local minima), named metabasin, followed by a sharp crossing to a different metabasin. 
The characteristic time between metabasin transitions is comparable to the structural relaxation time, suggesting that these transitions are relevant for the long time dynamics.  The crossing between metabasins is accompanied by very rapid diffusional jumps of several groups of dynamically correlated  particles. These particles form relatively compact clusters and act as cooperative relaxing units  responsible for the density relaxation. We find that these mobile particles are often characterized by an average coordination larger than four, i.e. are located in regions where the tetrahedral hydrogen bond network is distorted. \end{abstract}


\maketitle

\section{Introduction}

The elucidation of the physical underpinnings of glassy dynamics represents a central open problem in condensed matter~\cite{1,2,3}.
 Within this realm, the discovery that glassy systems are dynamically heterogeneous represented a major breakthrough and reinforced the interest in the Adam \& Gibbs heterogeneous scenario for glassy relaxation: the existence of  cooperatively relaxing regions in the sample whose timescales and sizes grow as the temperature is decreased in the supercooled regime~\cite{4,5,6,7,8,9}. 
A complimentary and powerful framework for this field has been provided by the so-called landscape paradigm~\cite{3,10,11,12,JSTAT}: 
the dynamics of a glassy system can be followed as the exploration it performs of its Potential Energy Landscape (PEL), the surface 
generated by the potential energy of the system as a function of particle coordinates. The PEL 
  can be partitioned into disjoint basins, where a  basin is unambiguously defined as the 
 set of points in configuration space connected to the same local minimum  (called    
inherent structure, IS) via a steepest-descendant  trajectory.  Closely related IS are arranged in metabasins (MB).
 At low enough temperature the dynamics of the system  can be decomposed in fast vibrations
 within IS and  structural rearrangement transitions between different IS. 
In this terms, the dynamics of the system can be studied following its dynamics between IS.
If the temperature is very low  the system doesn't have enough energy to easily overcome barriers
between PEL valleys,  this resulting in a  slowing down of the dynamics.
The link between these two descriptions has been already studied by means of molecular dynamics simulations~\cite{8,12,13,14,15}.
 An interesting finding in such direction for binary Lennard-Jones systems suggested a compelling picture
 for glassy dynamics~\cite{8}: the $\alpha$-relaxation corresponds to a small number of fast crossings from one metabasin to a
 neighboring one involving the collective motion of a significant number of particles that form a relatively compact cluster, called ``democratic cluster''
 or d-cluster. Since the influence of the structure on dynamics is only local in time~\cite{16}, the long time dynamics could be described as a 
random walk on metabasins~\cite{13,16} triggered by d-clusters~\cite{16}. Thus, it would be of great interest to learn whether these results hold 
valid for other glassy systems beyond binary Lennard-Jones systems so as to determine the generality of such a scenario. In this sense, the aim of the 
present work is to study the possible metabasin structure of supercooled water and its connections to local structure and dynamics. 

\section{Metabasin structure in supercooled water}

Water is not only important from its condition of being our most familiar liquid. Supercooled water exhibits an unusual behavior that has fostered 
intense experimental and theoretical work both in thermodynamics and dynamics in the past decades~\cite{17,18,19,20}. From a dynamical viewpoint,
 it has been shown computationally that supercooled water confirms the general picture of dynamical heterogeneities~\cite{21,14,9,poole}, 
with mobile molecules arranged in clusters usually linked by hydrogen bonds. The properties of its PEL have also been studied, including the characterization of transitions between ISs~\cite{14}. Within this context, it has also been demonstrated that the presence of molecules of high coordination (more than the usual four first neighbors demanded by its local, long-range disordered, tetrahedral structure) and of bifurcated hydrogen bonds (hydrogen bonds that link one hydrogen to two oxygens, instead of only one) enhance the local mobility~\cite{22,23,14}. That regions of greater local density and the presence of bifurcated hydrogen bonds aid molecular mobility is not surprising since the possibility for higher coordination and hydrogen bonding can help lowering the barrier to restructure  the hydrogen bond network. Indeed,for such water molecules, a new hydrogen bond can begin to be formed at the same time as the old one is being broken, thus lowering the energetic cost~\cite{14}.  
   
\begin{figure}[tb]
\includegraphics[width=0.9\linewidth]{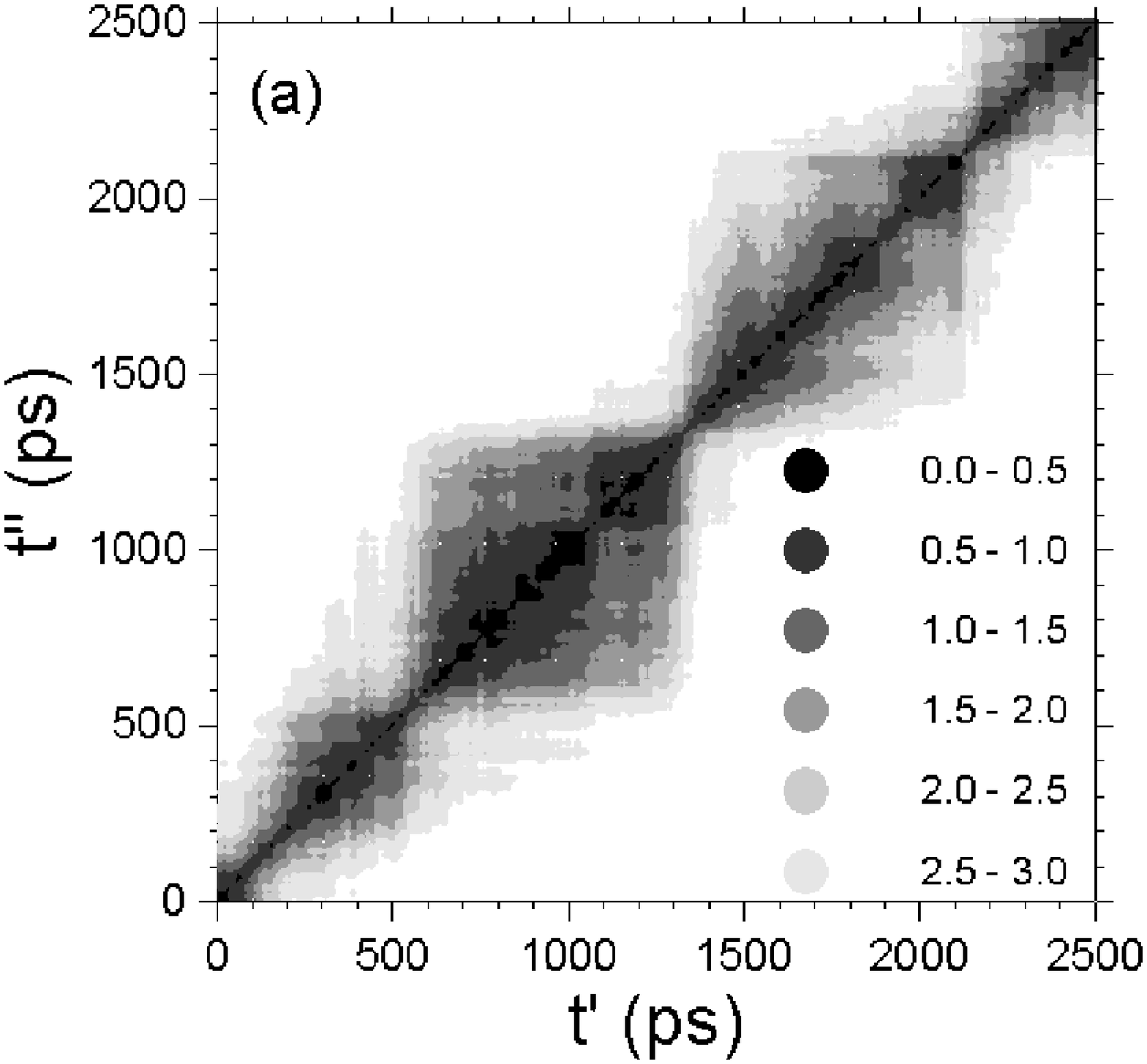}
\includegraphics[width=0.9\linewidth]{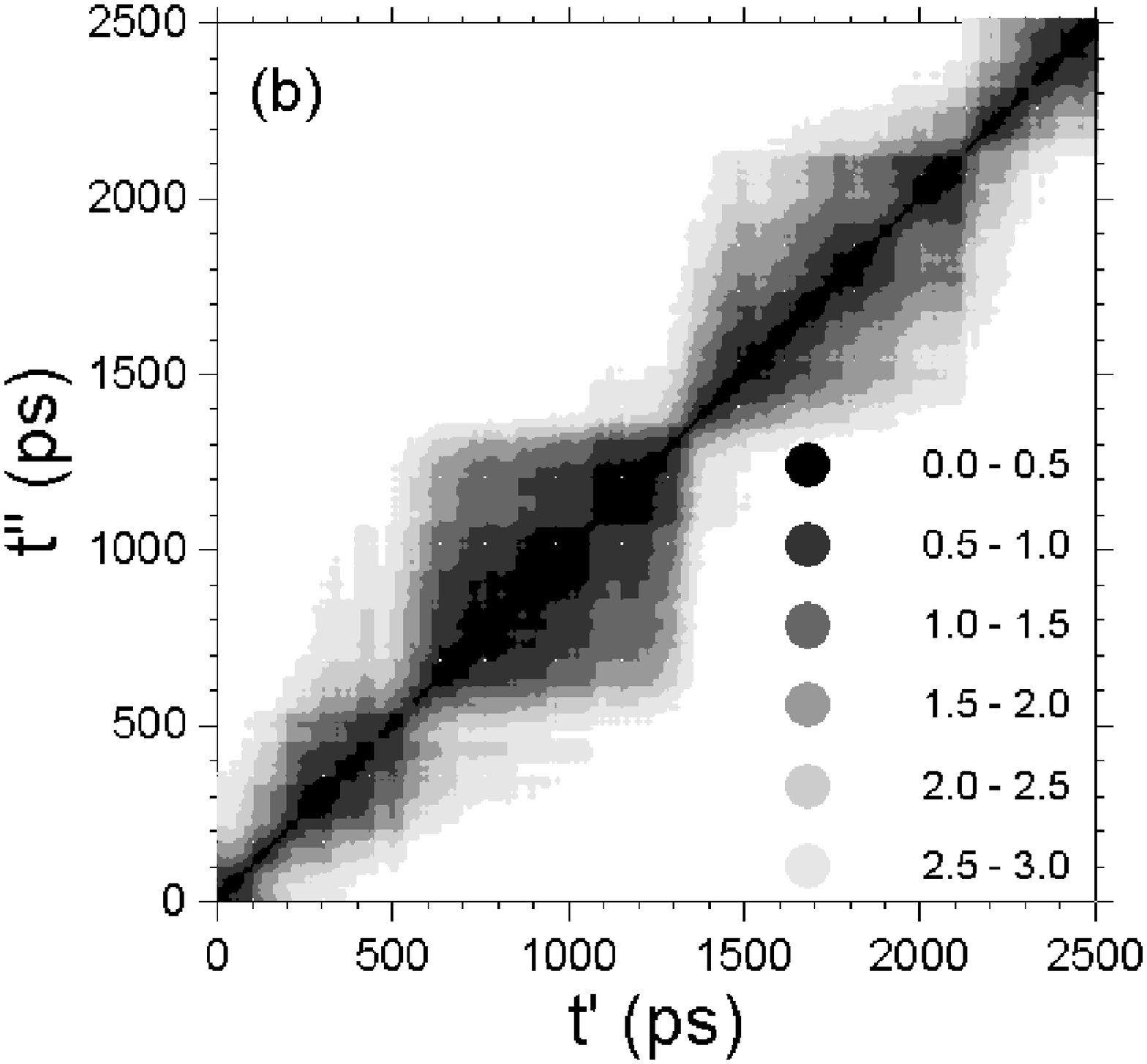}
\includegraphics[width=0.9\linewidth]{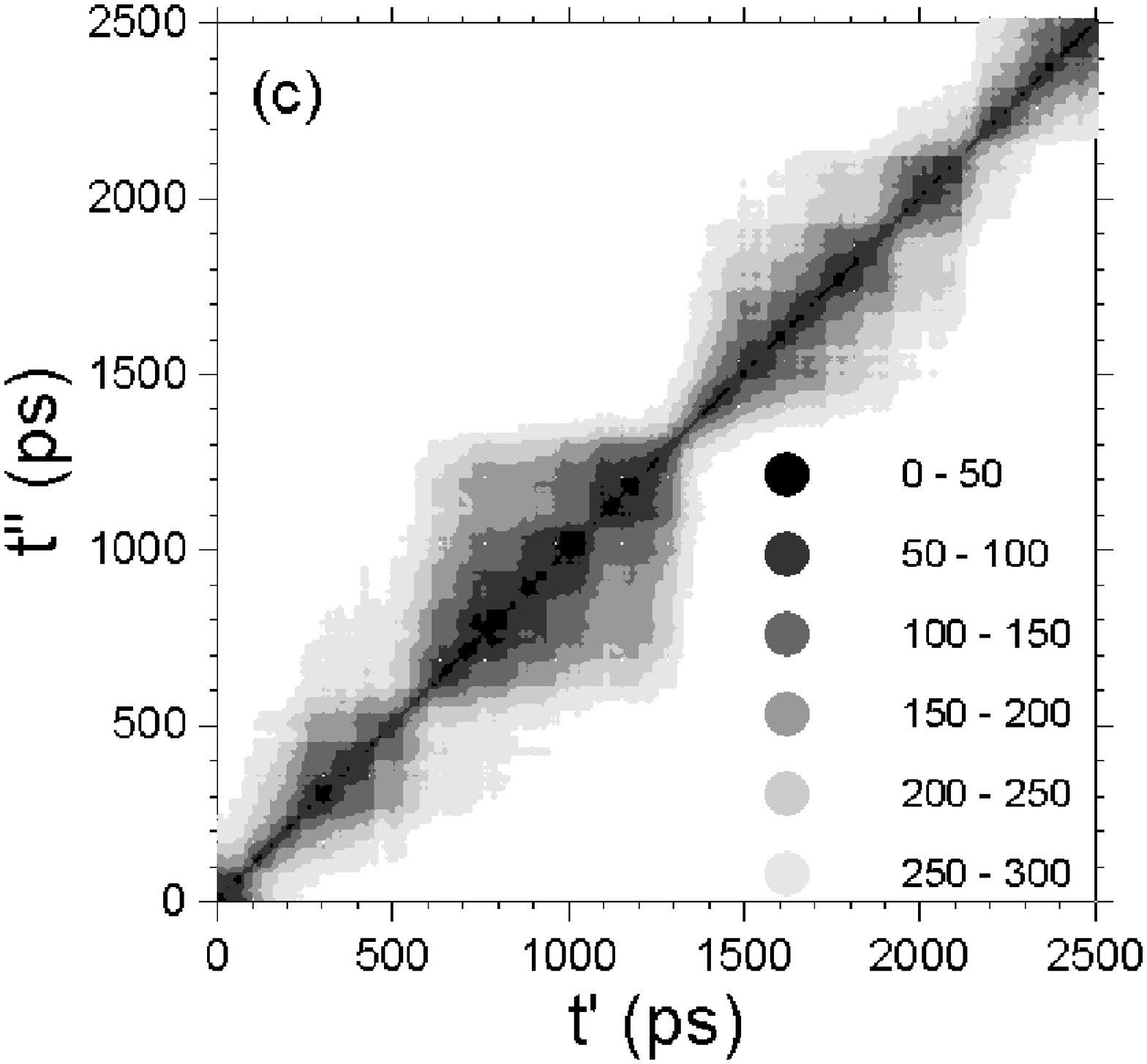}
\caption{(a) and (b): Distance matrix $\Delta^2(t',t'')$ respectively for real and inherent dynamics. (c): Hydrogen bond Hamming distance
 matrix of the system for $T=210$ K. The gray level correspond to values
 of $\Delta^2(t',t'')$ that are given to the right of the figures (units are \AA$^2$ for 
(a) and (b) and hydrogen bond changes for (c))}
\label{fig1}
\end{figure}

In this work we have studied a system of 216 water molecules interacting via the  simple point charge 
extended (SPC/E) model \cite{spce}   in the (NVE) ensemble.  The integration step is 1 $ft$ second, 
and long range interactions are taken in account using the reaction  field method.
At temperatures within the supercooled regime we investigated both the real dynamics (the instantaneous MD configurations) and the inherent 
dynamics, the dynamics of the quenched structures, which have been calculated performing periodical quenches of configurations obtained in the real trajectories 
  using a standard conjugate gradient minimization algorithm with $10^{-15}$ tolerance \cite{numericalrec}.
In order to study the possible metabasin structure of this system we
recorded 250 configurations at fixed time intervals of 10 ps --- at
density 1 g/cm$^3$ and T=210 K  ---. The total run time is larger than
the $\alpha$-relaxation. For these conditions the
$\alpha$-relaxation time has been estimated (as the time when the
intermediate incoherent scattering function, measured at the wave
vector of the first peak in the structure factor, has decayed to 1/e)
to be $\tau_\alpha=769 ps$ \cite{starr99}, both for the real and the
inherent dynamics. We shall present results for a run at $T=210$ K
but the same behavior was found for other trajectories at such
temperature and at other close temperatures. With the recorded
configurations we built the following distance matrix~\cite{8,24}, DM:
$\Delta^2(t',t'') = \frac{1}{N}\sum_{i=1}^N |{\bf r}_i(t')-{\bf
r}_i(t'')|^2$, where ${\bf r}_i(t)$ is the position of the oxygen of
molecule $i$ at time $t$ (in the real or the inherent
configuration). Thus $\Delta^2(t',t'')$ gives the system averaged
squared displacement of a molecule in the time interval that starts at
$t'$ and ends at $t''$. In other words, this distance matrix contains
the averaged squared distances between (the oxygens of) each recorded
configuration and all the other ones. For this kind of study, we must
investigate small systems, since for large systems different
subsystems would produce interfering results~\cite{8,13}. Thus, we
used $N=216$ molecules (however, the same qualitative results would be
obtained for a small subsystem immersed in a big system, thus ruling
out the possibility for finite size effects, as found for binary
Lennard-Jones systems~\cite{8}). For this small-sized system it has been
shown that  no significant finite-size effects are present in the
studied temperature range~\cite{starr99}.
. Fig.~\ref{fig1} shows typical results for a run at $T=210$ K. The gray level of the squares in the DM depicts the distance between the corresponding configurations, the darker the shading indicating the lower the distance between them. Fig.~\ref{fig1}a shows the real dynamics case while Fig.~\ref{fig1}b shows the inherent dynamics one. 
 Additionally, in Fig.~\ref{fig1}c we plot the hydrogen bonds Hamming distance matrix, that is, we calculated the elements of this matrix as the number of hydrogen bond changes between the corresponding structures at times $t'$ and $t''$ (in the inherent dynamics). For the identification of hydrogen bonds, HB, we employed a geometrical criterion consistent with previous mixed geometric, energetic and dynamic strategies~\cite{25}: two water molecules are linked by a HB if the O-O distance is lower than 0.35 nm (a distance that falls beyond the first peak of the O-O radial distribution function) and the O-H...O angle is lower than $60^\text{o}$. Nevertheless, the vast majority of linear HBs found for the inherent structures presented O-O distances lower than 0.3 nm and very low angles (angles of $40^\text{o}$ - $60^\text{o}$ occurred only for one of the bonds of bifurcated HBs, as we shall see later on). 
 Direct inspection of these graphs shows that the results of the real and the inherent dynamics are practically identical (thus, from now on, we shall refer further results to the inherent dynamics, unless indication on the contrary). From the island structure of these matrices a clear MB structure of the landscape is evident. That is, islands are made up of closely related or similar configurations (low $\Delta^2$) which are separated from the configurations of other islands by large distances.  From the size of the islands found for many different independent runs at $T=210K$ we can estimate the typical residence time in the MBs for this temperature to be close to the $\alpha$-relaxation time, $\tau_\alpha$. Given the small system size we expect this to be a good estimate (however, this timescale clearly depends on system size, since for a large system different subsystems would be undergoing MB transition events at different times). We also studied the case of $T=220K$ and $T=230K$ (always for systems of $216$ molecules) for which $\tau_\alpha$ is respectively around $8$ and $60$ times lower than that for $T=210K$ and found that for both cases the typical MB residence time is also close to the corresponding value of $\tau_\alpha$. Thus, $\tau_\alpha$ and the MB residence time seem to follow the same temperature dependence. The transitions between the MBs (which, as can be learnt from Fig.~\ref{fig1}, last typically a few tens of picoseconds) are fast events compared to the times for the exploration of the MBs. Thus, consistent with the situation for Lennard-Jones systems~\cite{8}, the MB transitions are the events that trigger the $\alpha$-relaxation in supercooled water. The hydrogen bonds Hamming DM showed in Fig.~\ref{fig1}c, which is very similar to the DMs of Fig.~\ref{fig1}a and b, makes evident the fact that MB transitions imply significant restructuring of the hydrogen bonds (HB) network of supercooled water.

\section{Metabasin dynamics and molecular mobility}

\begin{figure}[tb]
\vspace*{4mm}
\includegraphics[width=0.9\linewidth]{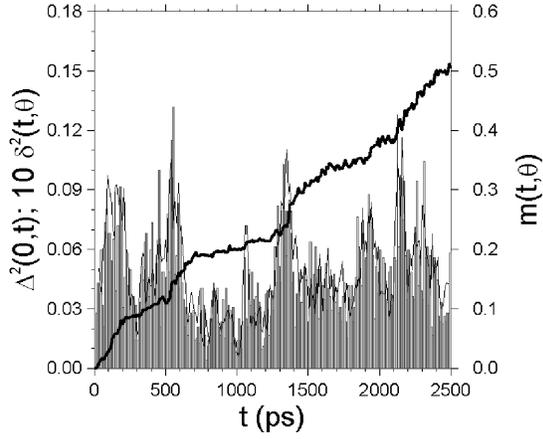}
\caption{Left scale: Average squared displacement $\Delta^2(0,t)$ (thick solid line) and $\delta^2(t,\theta)$ (thin solid line) for the trajectory given in Fig.~\ref{fig1}b. The value of $\theta$ is 40 ps for this last curve. For both curves the units are nm$^2$. Vertical bars (right scale): The function $m(t,\theta)$ which gives the fraction of oxygen atoms that moved more than the threshold value $r_{\rm th}= 0.06$ nm in the time interval $[t,t+\theta]$, using $\theta=10$ ps.}
\label{fig2}
\end{figure}

\begin{figure}[tb]
\includegraphics[width=0.9\linewidth]{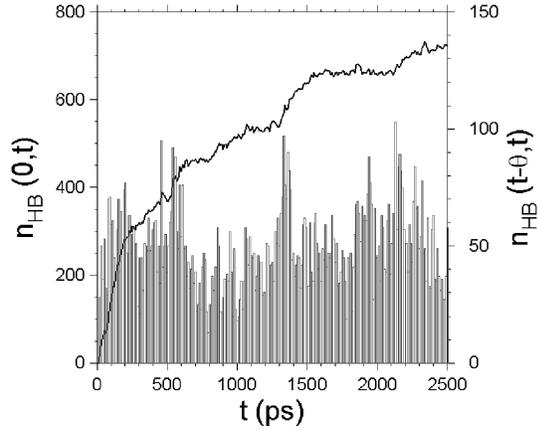}
\caption{Solid line (left scale): Hydrogen bonds (number) changes between a configuration at time t and the first configuration. Vertical bars (right scale): Hydrogen bonds (number) that have changed between successive recorded configurations (separated by a time interval of length 10 ps).} 
\label{fig3}
\end{figure}

\begin{figure}[tb]
\includegraphics[width=0.9\linewidth]{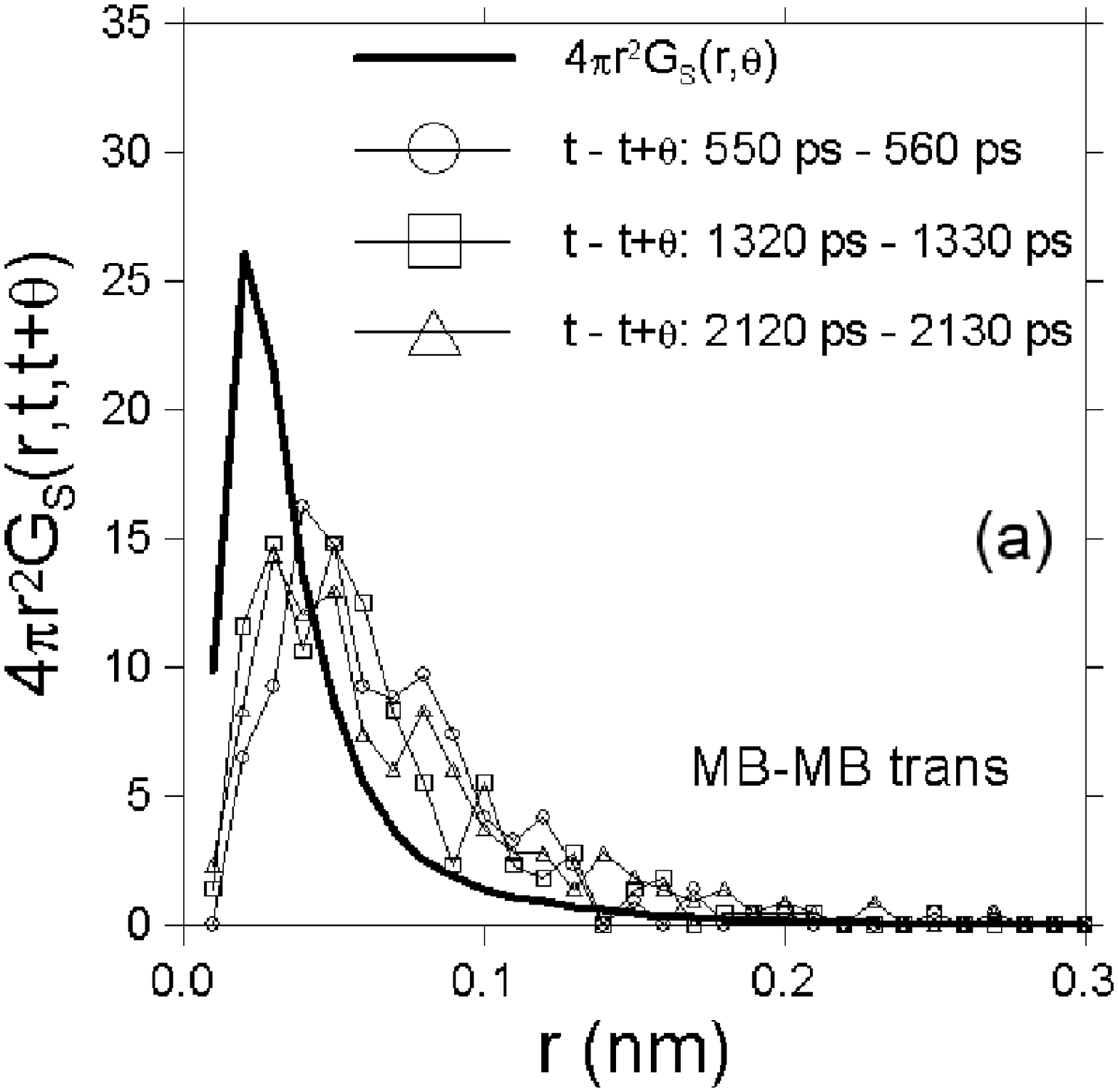}
\includegraphics[width=0.9\linewidth]{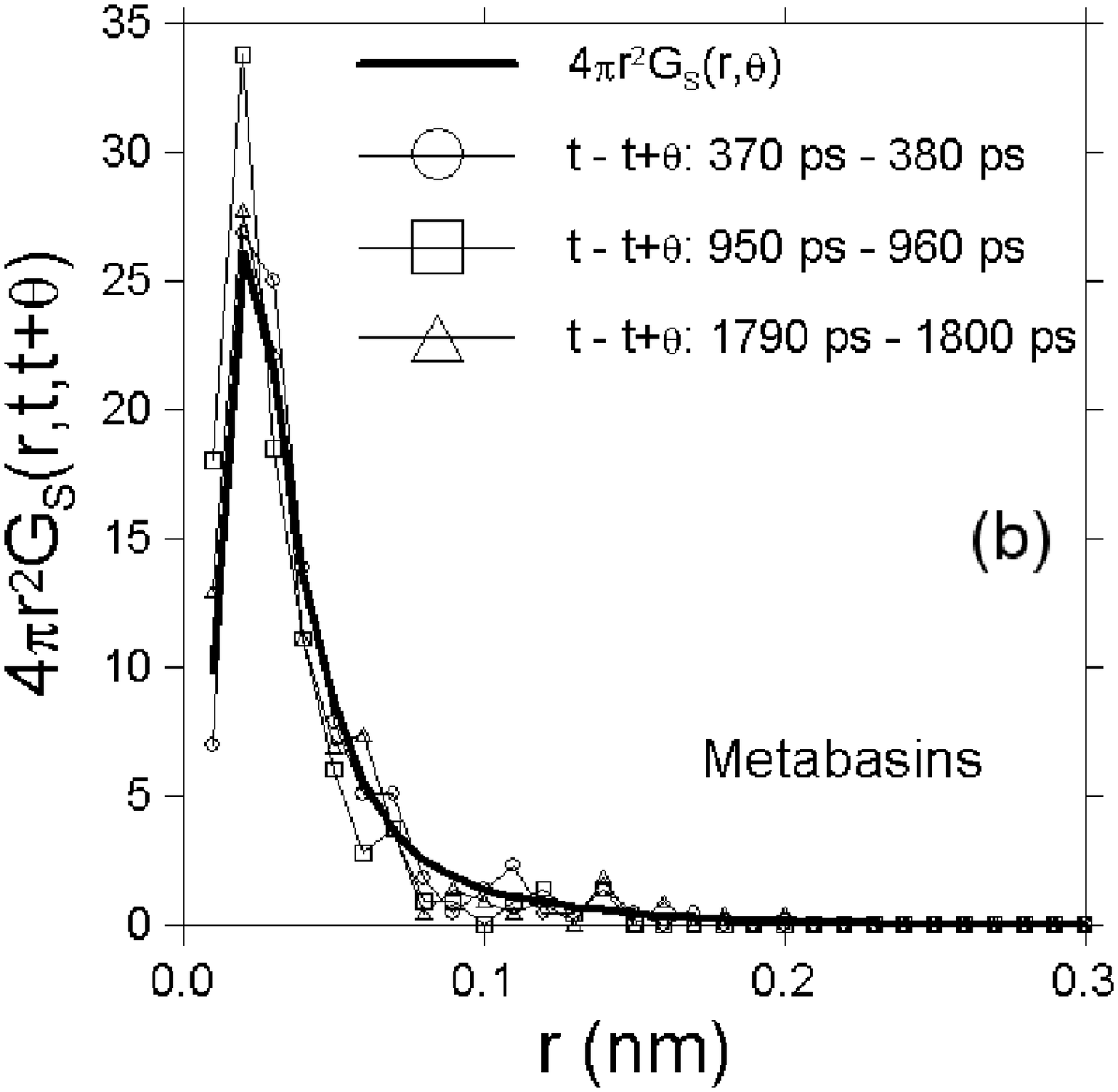}
\caption{The distribution function $\widehat{G}_s(r,t,t+\theta)$ for different
values of $t$ (curves with symbols). The value of $\theta$ is 10 ps. The
bold curve is $4\pi r^2 G_s(r,\theta)$, the self part of the van Hove function. a) Values of $t$ for which the system
is about to leave a MB. b) Values of $t$ in which the system is inside a MB.}
\label{fig4}
\end{figure}

A still open question is if  MB transitions are due to the large displacement of a few water molecules or to the result of more extensive collective rearrangements, as that found in binary Lennard-Jones systems~\cite{8}. To answer this point we performed a detailed analysis of molecular mobility, as shown in Fig.~\ref{fig2}. The thick solid line of such figure represents the average squared displacement plot $\Delta^2(0, t)$, that is, the first row of the DM of Fig.~\ref{fig1}(a) ~\cite{8}. The thin solid line depicts the average squared displacement of the molecules (by monitoring the displacements of the oxygen atoms) in time interval $\theta$=40 ps, $\delta^2(t,\theta)$, which corresponds to $\Delta^2(t',t'')$ measured along the diagonal $t''=t'+\theta$ ~\cite{8}. The bars of Fig.~\ref{fig2} display the function $m(t,\theta)$ ~\cite{8}, which represents the fraction of molecules that have moved more than a threshold of 0.06 nm in time intervals $[t,t+\theta]$ with $\theta$=10 ps (this threshold value is arbitrary but the reason for its choice will be given later on, and similar results arise for other values). From the thick solid line of Fig.~\ref{fig2} we can learn that MB transitions represent abrupt jumps (of 5-10 times the value of the squared distances between contiguous configurations separated by 10 ps). In turn, the thin solid line makes clear the fact that MB transitions imply a significant enhancement of molecular mobility, while the bars relate this enhancement of global mobility to the fraction of molecules moving. Thus, in agreement with previous findings for binary Lennard-Jones systems~\cite{8}, MB transitions imply the rearrangement of a substantial amount of the molecules of the system. That is, our results clearly demonstrate that also for supercooled water the $\alpha$-relaxation time corresponds to a small number of crossings from one metabasin to a neighboring one, each crossing being very rapid and involving the collective motion of a great number of particles.  From direct inspection of the function $m(t,\theta)$ in Fig.~\ref{fig2} we can see that in the MB transitions $35-45$\% of the particles move more than $0.06nm$ in $10 ps$, a time span slightly larger than $1$\% $\tau_\alpha$ (for comparison, an integration of the self part of the van Hove function, $4\pi r^2 G_s(r,\theta)$, from $0.06$ to infinity gives less than $0.15$, cf. Fig.~\ref{fig4}). The events that constitute the local exploration of the MB represent the plateau regions of the thick solid line of Fig.~\ref{fig2} and, while relevant to the $\beta$-relaxation, do not contribute to the structural relaxation of the system. Thus, these results demonstrate the prevailing role of metabasin transitions for the long time dynamics.
Fig.~\ref{fig3} solid line depicts $n_{HB}(0,t)$,the (number) difference in HBs between a configuration at time $t$ and the first configuration, while the bars indicate the function $n_{HB}(t,t+\theta)$, the number of HB that have changed between successive recorded configurations (separated by a time interval of length $\theta$). The similarity between this figure and Fig.~\ref{fig2} implies that the HB can also be used to signal MB transitions since such events entail a major rearrangement of the hydrogen bond network of the system.
In Fig.~\ref{fig4} we study the range of mobility of the molecules associated with MB transitions. The plot of Fig.~\ref{fig4}a corresponds to the Van Hove function $4\pi r^2 G_s(r,t,t+\theta)$ ~\cite{8} evaluated at chosen MB transition events (within $[t,t+\theta]$ time intervals, $\theta$=10 ps) of the run under study. In other words, Fig.~\ref{fig4}a displays the distribution of mobility for the water molecules (by looking at the oxygens) at such events. The solid line depicts the distribution function of oxygen displacements averaged over the whole run, that is, the van Hove function $4\pi r^2 G_s(r,\theta)$ (which gives the probability for an oxygen to be located at a distance $r$ from the origin after time $\theta$).
 Fig.~\ref{fig4}b is the analog of Fig.~\ref{fig4}a but for the function $4\pi r^2 G_s(r,t,t+\theta)$ evaluated at selected time intervals within MBs.
 From these plots we can learn that molecular motion is greatly enhanced at the
 MB transitions since the displacement distribution is not shifted compared to the van Hove function in Fig.~\ref{fig4}b and it is  significantly shifted to the right in Fig.~\ref{fig4}a.  Additionally, we can see that there is no clear preference for displacements of any given length. On average, the distribution functions for the MB transitions exceed the van Hove function at lengths $r$ around 0.025 nm - 0.03 nm. Thus, we could select mobile molecules as those whose displacement is larger than such cut off. If we had used such value for the bars of Fig.~\ref{fig2} we should have obtained the same qualitative shape. However, we used a larger value (0.06 nm) since this will be more practical for the following section where we shall display and study the clusters of mobile particles at MB transitions. We also calculated the same quantities shown in Fig.~\ref{fig4}a for the real dynamics --- as opposed to the IS dynamics --- where a similar behavior is found but with both kind of functions displaced to the right due to the thermal motion. These data are not reported in this manuscript.  While both approaches produce clear results, these are neater for the inherent dynamics.

\section{Democratic clusters and local water structure}

\begin{figure}[tb]
\includegraphics[width=0.9\linewidth]{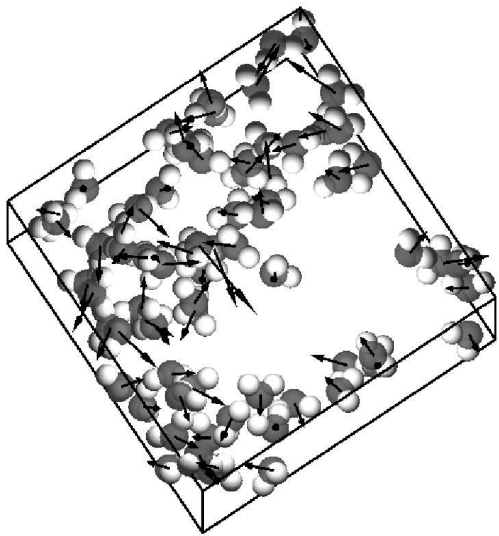}
\caption{Configuration snapshot of mobile molecules occurring in the MB-MB
transition $t=1320$ ps to $t=1330$ ps. The spheres (light and dark for the hydrogen and oxygen atoms, respectively ) give the location of the atoms of the water molecule before the rearrangement and the arrows point to the position of the oxygen atom after
the transition (to help visualize the particle movments, these arrows are twice the length of the corresponding displacement vectors.}
\label{fig5}
\end{figure}

\begin{figure}[tb]
\includegraphics[width=0.9\linewidth]{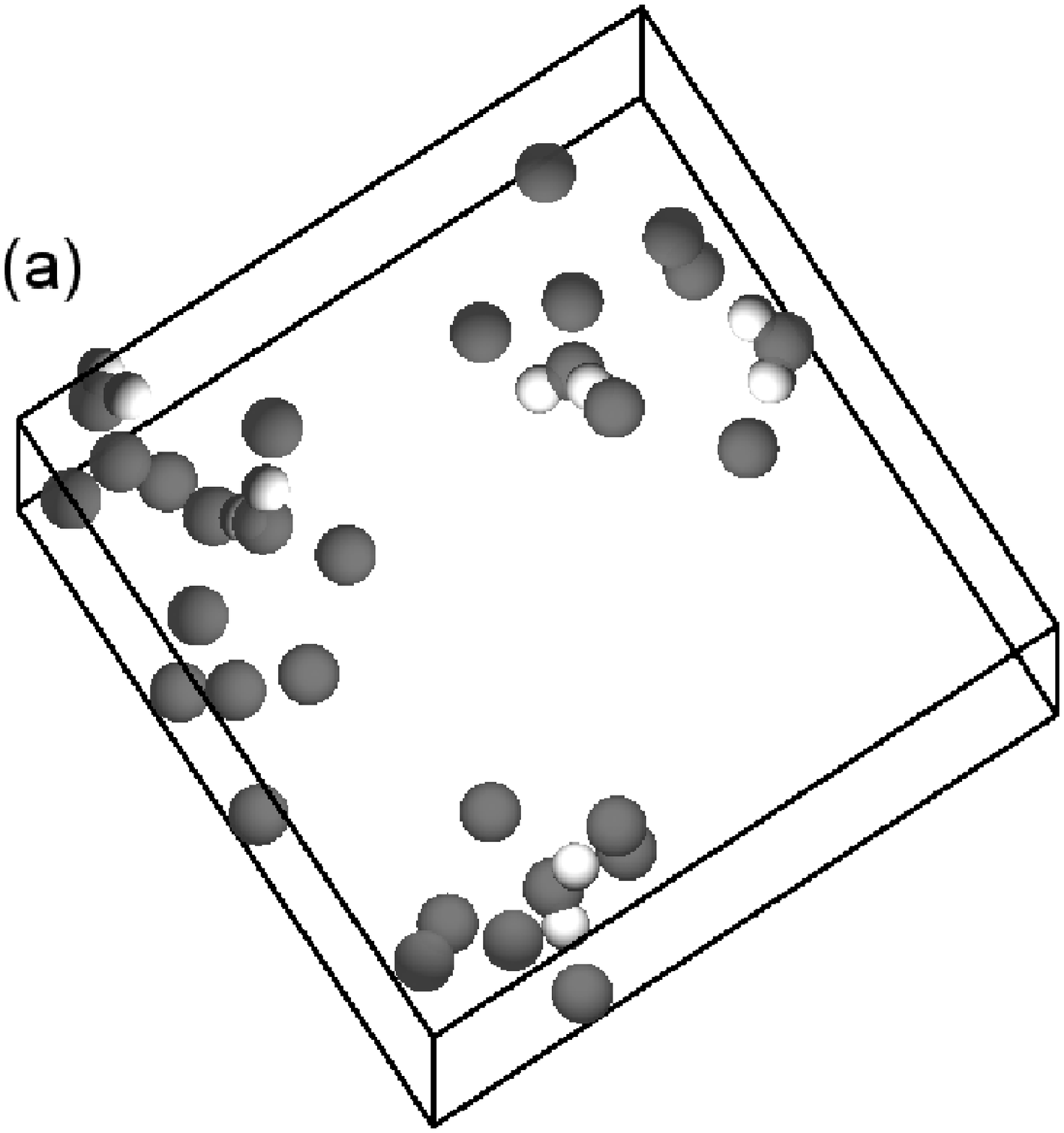}
\includegraphics[width=0.9\linewidth]{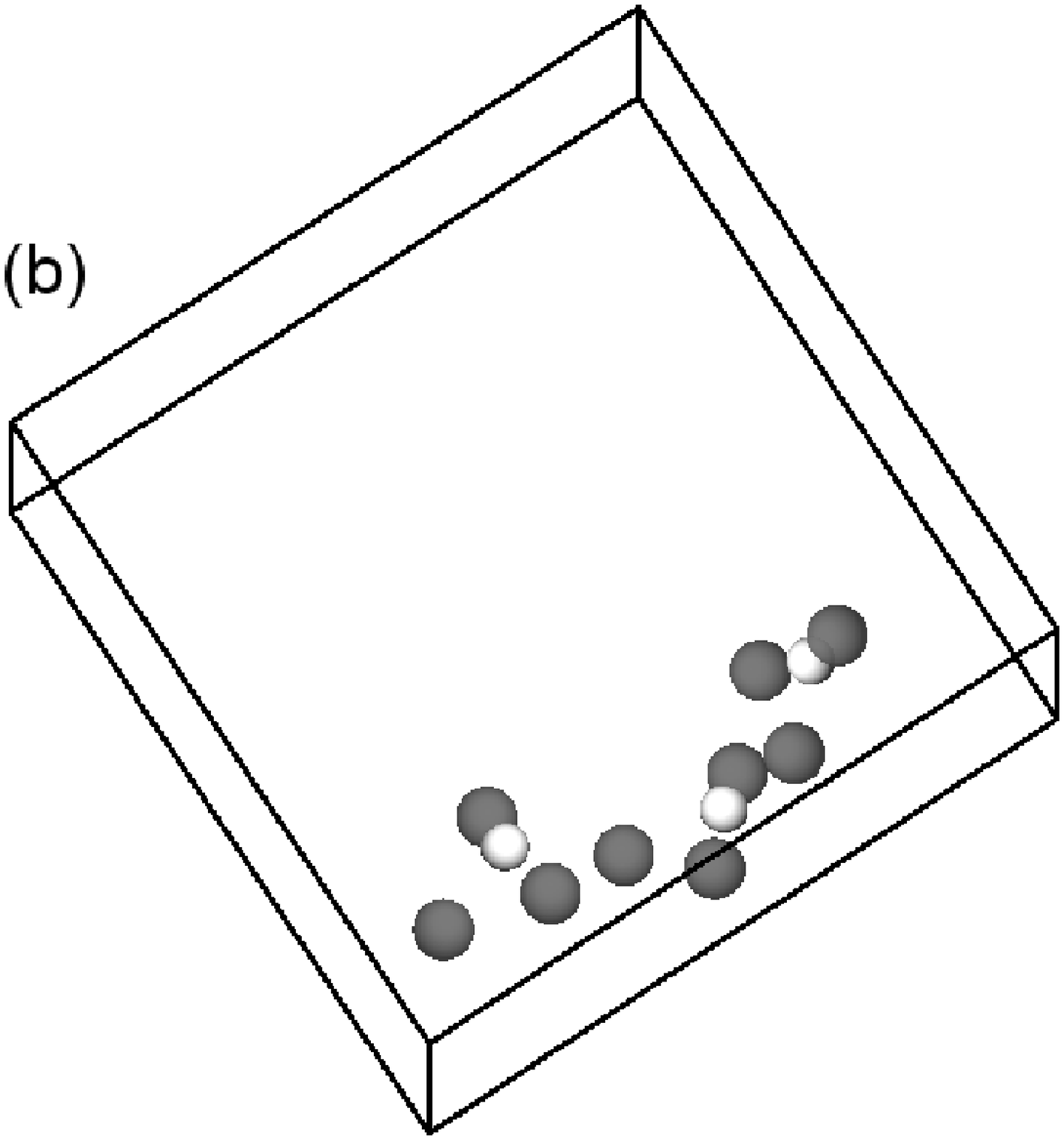}
\caption{Configuration snapshot of (a): molecules having more than four first neighbors within a distance of 0.30 nm together with such neighbors (the central water molecule is fully depicted with its corresponding hydrogen atoms, while only the oxygens of the first neighbors are included); (b): molecules with bifurcated hydrogen bonds, for the structure at time $t=1320$ ps. Only the oxygens of the corresponding molecules are indicated together with the hydrogens involved in the bifurcated bonds. Light and dark spheres are for the hydrogen and oxygen atoms, respectively.}
\label{fig6}
\end{figure}

\begin{figure}[tb]
\includegraphics[width=0.9\linewidth]{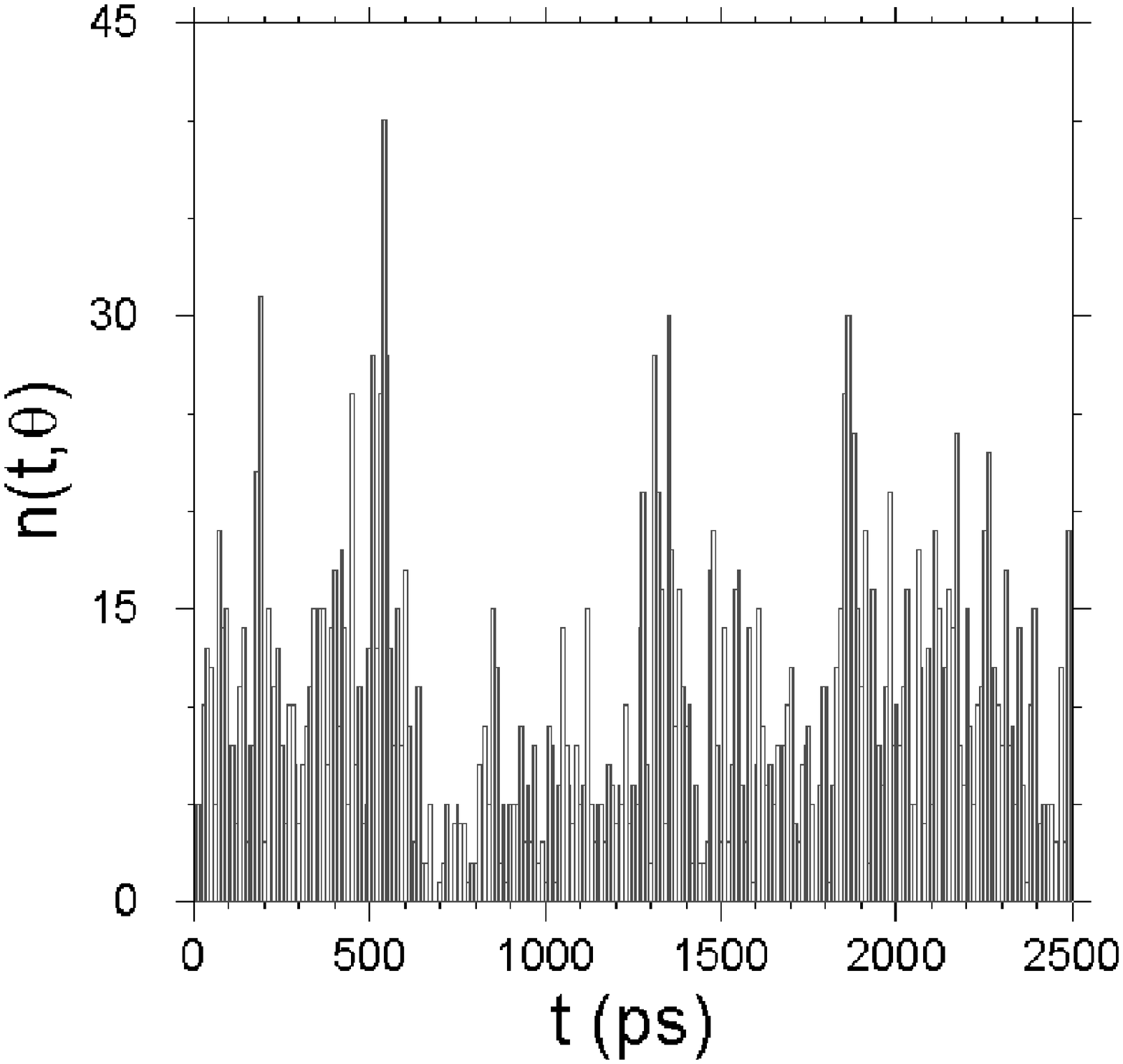}
\caption{The function $n(t,\theta)$ shows the number of coincidences or matches between the defects at time $t$ and the mobile molecules (with mobility greater than 0.06 nm) in the time interval $[t,t+\theta]$ , with $\theta=10$ ps.}
\label{fig7}
\end{figure}

Having established the main role of large scale molecular rearrangements to the MB dynamics and thus, on the long time dynamics of supercooled water, the aim of this section is to study the nature of such molecular motions. Thus, the spatial distribution of mobile molecules and its relation to the local structure of the system will be explored. We shall see that, consistent with the situation for binary Lennard-Jones systems~\cite{8}, relatively compact clusters of molecules conforming cooperative relaxing units emerge as responsible for the structural relaxation of the system. Moreover, we shall make evident the connection of these clusters with local structural ``defects'' such as highly coordinated water molecules and bifurcated hydrogen bonds, effects that have previously been regarded as mobility promoters~\cite{14,22,23}.
Fig.~\ref{fig5} shows a three-dimensional plot of the mobile particles (mobility greater than 0.06 nm) for one of the time intervals of Fig.~\ref{fig4}, namely that which goes from $t$=1320 ps to $t$=1330 ps, but similar results were found for other MB transitions. We show each mobile particle (oxygens and hydrogens) indicating the position at $t$=1320 ps and attaching a vector that shows its displacement to the position occupied at time $t$=1330 ps. A clear cluster arrangement of the mobile water molecules is evident from such picture. In this graph we can see that the mobile molecules are located close to the borders of the simulation box and an empty space occurs at center (the molecules in such region are not mobile). These kind of compact clusters resemble the situation in Lennard-Jones systems where were termed as ``democratic clusters'' or d-clusters~\cite{8}. These d-clusters represent potential candidates for the cooperatively relaxing regions proposed long ago by Adam \& Gibbs whose verification has remained elusive till now (a d-cluster would represents a event when a subsystem, that reached a state that permitted a rearrangement, has performed such transition, that is, it would constitute an active subsystem at such time). However, we note that no definite connection between the d-clusters and the CRRs can be done without an exhaustive study of the dependence of the size of the d-clusters with temperature and its relation to the structural relaxation time. The motions of many of the molecules of the d-clusters seem to be highly coordinated, with clear coherent flows of groups of molecules  (they move in an organized fashion, that is, carrying a common course, as can be verified by looking at the vectors attached to each atom in the 3D plot of the d-cluster of Fig.~\ref{fig5}).
To establish possible links between local structure and dynamics, we
studied the local environment of the mobile molecules that comprise
the d-clusters. An interesting connection between structure and
dynamics has been found previously for supercooled water~\cite{14} and
has been shown to be related to transitions between contiguous ISs of the PEL which entail the movement of a group of a few water molecules: mobility is facilitated by the presence of defects like highly coordinated molecules (molecules with more than four first neighbors) and molecules involved in bifurcated hydrogen bonds (molecules which are hydrogen bonded to two other water molecules via the same hydrogen). These regions of higher local density aid mobility by lowering the cost of hydrogen bond reformulation since the breakage of the old hydrogen bond implied in the motion of the mobile molecule is somehow counterbalanced by the simultaneous formation of the new one (a fact which would be otherwise anti-intuitive)~\cite{14}. Regarding bifurcated hydrogen bonds, we found that almost in all cases one of the bonds is almost linear (very low O-H...O angle) and short, while the other one is a bit longer and with a less favorable O-H...O angle between $40^\text{o}$ and $60^\text{o}$ (thus, this bond is clearly very weak). We note that minimized configurations present bifurcated HBs, a fact that speaks of certain stability of such ``species'' which can thus be regarded as intermediates rather than transition states for the interchange of hydrogen bonds. We shall show here results for the inherent dynamics, but the situation is similar for the real dynamics (albeit the 
identification of the  HBs is blurred  by thermal motion resulting in a larger 
number of bifurcated HBs). Fig. 6a shows the water molecules that posses more than four neighbors within a distance of 0.30 nm together with such neighbors, for the structure at time $t$=1320 ps (the first structure of the MB transition interval shown in Fig.~\ref{fig5}). In turn, Fig.~\ref{fig6}b shows the bifurcated HBs present at the same configuration. We can easily learn that both highly coordinated molecules and bifurcated HBs are located at the same regions of the mobile molecules of Fig.~\ref{fig5} while the central region of the box, where no mobile particles occur, is also empty. That is, a clear link exists between the regions of high coordination and bifurcated bonds in a structure previous to a MB transition event, and the region where the imminent d-cluster occurs, thus relating local structure to dynamics. For convenience, we shall term the water molecules that display coordination higher than four, the corresponding neighbors of such highly coordinated molecules and the molecules involved in bifurcated HBs as ``defects'' (for example, all the molecules displayed in Figs~\ref{fig6}a and b). We shall also define the function  $n(t,\theta)$ as the number of coincidences or matches between the defects of the configuration at time $t$ and the mobile molecules (with mobility greater than 0.06 nm) for the time interval $[t,t+\theta]$. The outcome of this quantity is shown in Fig.~\ref{fig7}. We can easily verify that this function is related to the MB transitions by noting the coincidences between the high bars of Fig.~\ref{fig7} and that of Fig.~\ref{fig2} (that indicates the previously defined function $m(t,\theta)$ which monitors the number of mobile molecules in the corresponding time interval). Thus, the function $n(t,\theta)$ can also be used to signal MB transitions.

\section{Conclusions}
 In this paper we have shown that 
 the dynamics of water in the supercooled region 
 consist in the exploration of a finite region of configuration space identified as metabasins
 followed by a sharp crossing to a different metabasin. The
 characteristic time between metabasin transitions is comparable to the structural relaxation time.
 These transitions are therefore relevant for the long time dynamics. 
The crossing between metabasins is accompanied by very rapid diffusional jumps of several
 groups of dynamically correlated  particles.
 These particles form relatively compact clusters
 and act as cooperative relaxing units  responsible for the density relaxation. 
These d-clusters represent potential candidates for the cooperatively relaxing regions proposed long ago by Adam \& Gibbs.
Finally, we related the mobility of the particles in the d-clusters to a studied quantity, namely the average local coordination number:
mobile particles are located  in regions where the tetrahedral hydrogen bond network is distorted.

\section{Acknowledgments}
J.A.R.F. thanks CONICET for a fellowship. G.A.A. is research fellow of CONICET and gratefully acknowledges financial support from ANPCyT, SeCyT and CONICET. ELN and FS are supported by MIUR-PRIN.

\begin{thebibliography}{10}

\bibitem{1}
C. A. Angell,
J. Non-Cryst. Solids 131, {\bf 13} (1991).

\bibitem{2}
M. D. Ediger,
Annu. Rev. Phys. Chem. {\bf 51}, 99 (2000).

\bibitem{3}
P. G. Debenedetti and F. H. Stillinger,
Nature {\bf 410}, 259 (2001).

\bibitem{4}
K. Schmidt-Rohr and H. W. Spiess,
Phys. Rev. Lett. {\bf 66}, 3020 (1991).

\bibitem{5}
M. T. Cicerone, F. R. Blackburn, and M. D. Ediger, 
J. Chem.  Phys. {\bf 102}, 471 (1995).

\bibitem{6}
C. Donati, J. F. Douglas, W. Kob, S. J. Plimpton, P. H. Poole and S. C. Glotzer,
Phys. Rev. Lett. {\bf 80}, 2338 (1998).

\bibitem{7}
M. Vogel, B. Doliwa, A. Heuer, and S. C. Glotzer,
J. Chem. Phys. {\bf 120}, 4404 (2004).

\bibitem{8}
G. A. Appignanesi, J. A. Rodr\'{\i}guez Fris, R. A. Montani, and W. Kob,
Phys. Rev. Lett. {\bf 96}, 057801 (2006).

\bibitem{9}
N. Giovambattista, S. V. Buldyrev, F. W. Starr and  H. E. Stanley,
Phys. Rev. Lett. {\bf 90}, 085506 (2003).

\bibitem{10}
C.A. Angell,
Nature {\bf 393}, 521 (1998).

\bibitem{11}
S. Sastry, P. G. Debenedetti and F. H. Stillinger,
Nature {\bf 393}, 554 (1998).

\bibitem{12}
G. A. Appignanesi and R. A. Montani,
J. Non-Cryst. Solids {\bf 337}, 109 (2004).

\bibitem{JSTAT}
F. Sciortino,
J. Stat. Mech. {\bf 050515}, (2005).

\bibitem{13}
B. Doliwa and A. Heuer,
Phys. Rev. E {\bf 67}, 031506 (2003). A. Heuer, B. Doliwa and A. Saksaengwijit,
Phys. Rev. E  {\bf 72} 021503 (2005).

\bibitem{14}
N. Giovambattista, F. W. Starr, F. Sciortino, S. V. Buldyrev and H. E. Stanley,
Phys. Rev. E {\bf 65}, 041502 (2002).

\bibitem{15}
T. B. Schr\"{o}der, S. Sastry, J. C. Dyre and S. C. Glotzer,
J. Chem. Phys. {\bf 112}, 9834 (2000).

\bibitem{16}
G. A. Appignanesi, J. A. Rodríguez Fris and M. A. Frechero,
Phys. Rev. Lett {\bf 96}, 237803 (2006).

\bibitem{17}
C. A. Angell,
Annu. Rev. Phys. Chem. {\bf 34}, 593 (1983).

\bibitem{18}
P. G. Debenedetti,
{\it Metastable Liquids}, Princeton University Press, Princeton, 1996.

\bibitem{19}
C. A. Angell,
Chem. Rev. {\bf 102}, 2627 (2002).

\bibitem{20}
C. A. Angell,
Annu. Rev. Phys. Chem. {\bf 55}, 559 (2004).

\bibitem{21}
F. Sciortino, P. Gallo, P. Tartaglia, and S.-H. Chen,
Phys. Rev. E {\bf 54}, 6331 (1996).

\bibitem{poole}
G. S. Matharoo, M.S. Gulam Razul and P. H. Poole,
 cond-mat/0607014 (2006).

\bibitem{22}
F. Sciortino, A. Geiger and H. E. Stanley,
J. Chem. Phys. {\bf 96}, 3857 (1992).

\bibitem{23}
F. Sciortino, A. Geiger and H. E. Stanley,
Phys. Rev. Lett. {\bf 65}, 3452 (1990).

\bibitem{spce}
H. J. C. Berendsen ,  J. R. Grigera and T. P. Stroatsma, 
{ J. Phys. Chem.} {\bf 91},  6269  (1987). 


\bibitem{numericalrec} W. H. Press, B. P. Flannery, S. A. Teukolsky and W. T. Vetterling {\it Numerical recepies}, Cambridge University Press, Cambridge, 1986.

\bibitem{starr99}
F. W. Starr, F. Sciortino and H. E. Stanley,
Phys. Rev. E {\bf 60}, 6757 (1999).

\bibitem{24}
I. Ohmine,
J. Phys. Chem. {\bf 99}, 6765 (1995).

\bibitem{25}
F. Sciortino and S.L. Fornili,
J.  Chem. Phys. {\bf 90}, 2786 (1989).

\end{thebibliography}
\end{document}